\documentclass{article}
\usepackage{amsmath, amssymb}
 \RequirePackage[colorlinks,citecolor=blue,linkcolor=blue,urlcolor=blue,pagebackref]{hyperref}

\newtheorem{lemma}{Lemma}
\newtheorem{cor}{Corollary}
\newtheorem{definition}{Definition}
\newtheorem{rem}{Remark}

\author{Maria Roginskaya}

\begin{document}

\title{Production and Just Distribution in Diverse Society}
\maketitle

\begin{abstract}
In this paper, we want to investigate dynamics of productivity in a society which is diverse when it comes to both the productivity and the perception of justice in distribution.    
\end{abstract}

\section{Introduction}
The economy has long focused on two main subjects of research: the {\bf production} and {\bf distribution} of resources within a society. These two topics have often been studied separately, even though it is evident that production directly influences distribution. Conversely, distribution is typically regarded as dependent on production, and not vice versa, although in reality, the relationship is far more reciprocal.

Production both depends on and affects a variety of physical and social factors, such as scientific knowledge and the social structures required for collaborative work.

Distribution, in turn, not only both defines and is defined by social structures, but is also intertwined with a wide range of human activities—including religion and literature—which serve to reinforce social norms supporting particular distributive arrangements. Through these socially imposed norms, members of a society are often compelled to accept the prevailing system of distribution. Frequently independent of individual preferences, a single distribution scheme has historically been upheld by religion, culture, and tradition as the just and proper way to allocate produced resources.

Today, we live in a society characterized by diversity. While differences in needs and abilities have always existed, now even the traditionally shared cultural aspects—such as modes of moral reasoning and perceptions of what is morally right—vary significantly. As a result, the concept of justice, once assumed to be universal, may differ from person to person, often leading to conflict and resentment.

The premises of the present paper can be summarized as follows:

\begin{itemize}
\item The total production of a society results from the combined efforts of its individual members, each of whom has a distinct productive capacity.
\item The distribution of produced resources follows a commonly recognized scheme, understood by all members of the society\footnote{The information assumption is a simplification and is not fully used}.
\item Each individual judges the fairness of the distribution scheme according to their own conception of justice, shaped by personal beliefs, knowledge, and cultural traditions.
\item The perceived fairness or unfairness of the distribution affects individual productivity—specifically, perceiving the system as just tends to increase productivity, while perceiving it as unjust tends to decrease it.
\end{itemize}

We now aim to construct a mathematical model that reflects these assumptions under significant simplifications. The models introduced describe economic development over time. We assume that the individual members of society produce some abstract “value,” and that the total value produced is the sum of individual contributions (this defines a productive society). At each stage, the total value is redistributed according to a predefined scheme. Simultaneously, each member holds a personal conception of what constitutes a “just” distribution. In each iteration of the process, an individual’s sense of being treated justly or unjustly influences their productivity in the following period. (This assumption aligns with a substantial body of research; see, for example, \cite{JPR}.)

The main objective of this paper is to formulate a simple taxation model for a diverse productive society and to analyze its behavior in elementary cases. Even this simplified model yields results that can be interpreted as reflecting real economic behaviors—such as the refusal to produce despite capacity, or the choice to produce less than one’s full potential—depending on one’s perception of what constitutes a “just” distribution.

\section{Definition of a Productive Society and Justice Schemes}

\begin{definition}
A {\bf productive society} is defined as a collection of $n$ individuals such that, at each discrete time period, every individual has a non-negative {\bf productive capacity}, {\bf actual productivity}, and adheres to a certain distribution scheme considered "just."

A {\bf distribution scheme} is a rule that determines how the total production of the society is allocated among its members, based on their individual contributions.
\end{definition}

We consider time in discrete intervals, corresponding to units such as accounting years.

Mathematically, at each time $t$, a productive society is represented by $n$ triples $(a_i, p_i, c_i)$, where $a_i, p_i \in \Bbb{R}_+$ denote the individual's capacity and actual productivity, respectively, and $c_i$ is a function that determines the individual's perception of a just distribution. This function depends on the individual's own productivity $p_i$ and the average productivity $P$ (we use average productivity instead of total productivity for simplicity).

\begin{definition}
At each time step, redistribution is performed according to a society-imposed distribution scheme $c(p_i, P)$. Each individual may consider this scheme to be just or unjust. Specifically, individual $i$ perceives the distribution as just if $c_i(p_i, P) \leq c(p_i, P)$ and unjust otherwise, i.e., if $c_i(p_i, P) > c(p_i, P)$.

The productivity of individual $i$ at time $t+1$ is then determined by a function of their capacity, current productivity, and their perception of justice, according to the rule:
\[
p_i(t+1) = J(a_i, p_i, c(p_i, P), c_i(p_i, P)),
\]
where $J$ is a suitable function defined on $\mathbb{R}^4$.
\end{definition}

Although the definition above allows for a broad range of behaviors, we will simplify the model by assuming that $a_i$, $c_i$, and $c$ remain constant over time. This assumption excludes possible changes in individual capacity due to aging, training, or technological advancements, as well as cultural or policy shifts in distribution norms.

\begin{rem}
We will consider a {\bf closed} productive society, meaning that the total production is fully redistributed: 
\[
\sum_i p_i = nP = \sum_i c(p_i, P).
\]
In real-world societies, additional value may come from sources such as natural resources, borrowing, or public liabilities. These factors may alter the total amount available for redistribution, but they are not considered in this model.
\end{rem}
\section{A Very Simple Example}

To illustrate the model with a simple example, we consider two distribution schemes: 
\begin{itemize}
    \item "Equal" (E): $c(p, P) = P$,  
    \item "Desert-Based" (D): $c(p, P) = p$.
\end{itemize}
If an individual feels they are being treated justly, they will increase their productivity within the limits of their capacity. Otherwise, they will decrease it.

A natural candidate for the \textbf{adjustment function} $J$ is a proportional increase or decrease in productivity (bounded by capacity), depending on whether the individual perceives the treatment as just or unjust. The model, on the other hand,  becomes more analytically tractable with the following definition:

\[
J(a_i, p_i, c, c_i) =
\begin{cases}
\min(a_i, p_i + (c - c_i)) & \text{if } c \geq c_i, \\
\max(0, p_i - (c_i - c)) & \text{if } c < c_i.
\end{cases}
\]

We consider a basic scenario with three individuals (A, B, C), where:
\begin{itemize}
    \item A can produce up to 1 unit of value,
    \item B can produce up to 2 units,
    \item C can produce up to 3 units.
\end{itemize}

In all examples below, at the initial step, each individual produces at their maximum capacity.

---

{\bf Case 0:} All individuals accept the same principle (E or D), and the distribution is made accordingly. In this case, each person considers the outcome just, and no adjustments are made in any subsequent time steps. The system remains stable.

This is shown in the tables below:

\begin{table}[h!]
\centering
\begin{minipage}{0.45\textwidth}
\centering
\begin{tabular}{cccc}
\ & A & B & C\\
$a_i$ & 1 & 2 & 3\\
E/D & E & E & E \\
$p_i(0)$ & 1 & 2 & 3 \\
$P(0)$ & 2 & 2 & 2 \\
$c_i(0)$ & 2 & 2 & 2\\
$c(0)$ & 2 & 2 & 2\\
$p_i(1)$ & 1 & 2 & 3\\
\vdots & \vdots & \vdots & \vdots \\
$p_i(n)$ & 1 & 2 & 3 \\
\end{tabular}

{Case 0: All accept E; E imposed}
\label{tab:E}
\end{minipage}
\hfill
\begin{minipage}{0.45\textwidth}
\centering
\begin{tabular}{cccc}
\ & A & B & C\\
$a_i$ & 1 & 2 & 3\\
E/D & D & D & D \\
$p_i(0)$ & 1 & 2 & 3 \\
$P(0)$ & 2 & 2 & 2 \\
$c_i(0)$ & 1 & 2 & 3\\
$c(0)$ & 1 & 2 & 3\\
$p_i(1)$ & 1 & 2 & 3\\
\vdots & \vdots & \vdots & \vdots \\
$p_i(n)$ & 1 & 2 & 3 \\
\end{tabular}

{Case 0: All accept D; D imposed}
\label{tab:D}
\end{minipage}
\end{table}

---

{\bf Case 1:} All individuals believe in D (desert-based distribution), but E (equal distribution) is imposed. 

After the first step, everyone receives 2 units. However, individual C believes they deserved 3 and therefore feels unjustly treated. In response, C reduces productivity to 2 in the next round. The new average becomes $P = 5/3$, and all receive $5/3$ in the second step. Now B and C both feel unjustly treated and reduce their productivity further, and so on.

The reader can verify that the system evolves as:
\[
(1,\ 1 + (2/3)^{t-1},\ 1 + (2/3)^{t-1}), \quad t \geq 2,
\]
converging to (but never exactly reaching) a stable state where all produce equally: $p_i = 1$.

\begin{table}[h!]
\centering
\begin{minipage}{0.45\textwidth}
\centering
\begin{tabular}{cccc}
\ & A & B & C\\
$a_i$ & 1 & 2 & 3\\
E/D & D & D & D \\
$p_i(0)$ & 1 & 2 & 3 \\
$P(0)$ & 2 & 2 & 2 \\
$c_i(0)$ & 1 & 2 & 3\\
$c(0)$ & 2 & 2 & 2\\
$p_i(1)$ & 1 & 2 & 2\\
$P(1)$ & 5/3 & 5/3 & 5/3  \\
$c_i(1)$ & 1 & 2 & 2 \\
$c(1)$ & 5/3 & 5/3 & 5/3 \\
$p_i(2)$ & 1 & 5/3 & 5/3  \\
\vdots & \vdots & \vdots & \vdots \\
$p_i(t)$ & 1 & $1 + (2/3)^{t-1}$ & $1+ (2/3)^{t-1}$ \\
\end{tabular}

{Case 1: All accept D; E imposed}
\end{minipage}
\hfill
\begin{minipage}{0.45\textwidth}
\centering
\begin{tabular}{cccc}
\ & A & B & C\\
$a_i$ & 1 & 2 & 3\\
E/D & E & E & E \\
$p_i(0)$ & 1 & 2 & 3 \\
$P(0)$ & 2 & 2 & 2 \\
$c_i(0)$ & 2 & 2 & 2\\
$c(0)$ & 1 & 2 & 3\\
$p_i(1)$ & 0 & 2 & 3\\
$P(1)$ & 5/3 & 5/3 & 5/3  \\
$c_i(1)$ & 5/3 & 5/3 & 5/3 \\
$c(1)$ & 0 & 2 & 3 \\
$p_i(2)$ & 0 & 2 & 3  \\
\vdots & \vdots & \vdots & \vdots \\
$p_i(t)$ & 0 & 2 & 3 \\
\end{tabular}

{Case 2: All accept E; D imposed}
\end{minipage}
\end{table}

---

{\bf Case 2:} All individuals believe in E (equal distribution), but D is applied. A feels unjustly treated (receiving 1 instead of the expected 2), and thus reduces productivity to 0 after the first iteration. B and C consider the distribution as just and maintain the same production. The system then stabilizes.

---

{\bf Case 3:} Two individuals accept E, and E is applied. The outcome depends on C's beliefs:
\begin{itemize}
    \item If C also accepts E, the system remains stable at maximum capacity.
    \item If C believes in D, then in the next round C reduces productivity to 2. This causes a drop in average productivity, and only C continues adjusting. The system converges to (but never reaches) a fixed point:  A produces 1, B produces 2, and C stabilizes at 1.5.
\end{itemize}
Details are left to the reader.

---

{\bf Case 4:} Two individuals accept D, and D is applied. If A is the one who accepts E, A feels under-compensated and reduces productivity to 0, while B and C continue to produce at maximal capacity. In all other scenarios, the system remains stable at full productivity.

---

We do not aim to exhaust all possible configurations, but this example serves to illustrate the fundamental dynamics of the model.

\section{Simple Taxation Model and Its Stable Points}

A slightly more complex model arises when the schemes of just distribution in use represent a mixture of schemes E and D from the previous section.

\begin{definition}
A \emph{distribution scheme} is said to be a \emph{taxation at level} $\alpha$, where \( \alpha \in [0,1] \), if
\[
c(p, P) = \alpha P + (1 - \alpha)p,
\]
where \( P \) denotes the average production of the group and \( p \) the production of an individual.
\end{definition}

This model corresponds to a flat taxation scheme, where the collective tax is redistributed equally without altering the total value of production. The previously discussed cases correspond to the extreme values \( \alpha = 0 \) and \( \alpha = 1 \).

The behavior of individuals at each iteration can be described as follows.

\begin{lemma}
If an individual accepts a level of justice in distribution given by \( \alpha_i \), while the imposed distribution operates at level \( \alpha \), then the following outcomes are possible:
\begin{enumerate}
    \item If \( \alpha_i > \alpha \) and \( p > P \): the individual perceives the distribution as just and either increases productivity or maintains full capacity.
    \item If \( \alpha_i < \alpha \) and \( p > P \): the individual perceives the distribution as unjust and reduces productivity.
    \item If \( \alpha_i > \alpha \) and \( p < P \): the individual perceives the distribution as unjust and reduces productivity.
    \item If \( \alpha_i < \alpha \) and \( p < P \): the individual perceives the distribution as just and either increases productivity or remains at full capacity.
    \item If $\alpha_i=\alpha$ or $p_i=P$: the individual perceives the distribution as just and does not change ones productivity.
\end{enumerate}
\end{lemma}

To simplify the discussion, let us refer to individuals with \( \alpha_i < \alpha \) as \emph{liberals}, and those with \( \alpha_i > \alpha \) as \emph{socialists}\footnote{The individuals with $\alpha_i=\alpha$ would not change the productivity, but can affect the total productivity. For simplicity we will assume that they do not exist.}. The following result describes the conditions under which the society reaches a stable equilibrium.

\begin{cor}
The model attains stability when the average production \( P \) is such that the above-average socialists produce at full capacity, thereby generating enough additional value to compensate for (i) socialists whose capacity is below \( P \), who in this state produce nothing, and (ii) liberals with capacity below \( P \), who produce at full capacity. A liberal whose capacity exceeds \( P \) produces a reduced amount equal to \( P \)\footnote{It may happen that a socialist produces exactly average and stay at this level of production, but this position is not a limiting point for a non-stable system.}.
\end{cor}

{\bf Proof of the Lemma}
Consider the actual outcome
\[
c(p_i, P) = \alpha P + (1 - \alpha)p_i,
\]
and the outcome the individual considers just:
\[
c_i(p_i, P) = \alpha_i P + (1 - \alpha_i)p_i.
\]
The difference between the two is
\[
c - c_i = (\alpha - \alpha_i)(P - p_i).
\]
This difference is positive when either:
\begin{itemize}
    \item \( \alpha > \alpha_i \) (the individual is liberal) and \( P > p_i \), or
    \item \( \alpha < \alpha_i \) (the individual is socialist) and \( P < p_i \).
\end{itemize}
In both cases, individuals perceive the outcome as just and increase productivity up to their capacity. When \( \alpha = \alpha_i \) or $P=p_i$, the individual considers the distribution just, yet the chosen function $J$ implies that productivity remains unchanged. In the remaining cases, they perceive it as unjust and reduce productivity.
\hfill $\Box$

In a productive system, the stable configuration is largely unaffected by liberals whose capacity exceeds the average, as they contribute exactly what they receive in distribution. Conversely, socialists either work at full capacity or not at all. The system’s average output \( P \) adjusts so that the socialists working at full capacity produce enough to offset the deficits of lower-capacity liberals (who work to their full capacity) and non-producing socialists.

\medskip
\noindent\textbf{Example.}
Consider five individuals with capacities 
\[
a_1 = 1, \quad a_2 = 2, \quad a_3 = 2, \quad a_4 = 4, \quad a_5 = 10,
\]
and parameters
\[
\alpha = 0.3, \quad 
\alpha_1 = 0.4, \quad 
\alpha_2 = 0.6, \quad 
\alpha_3 = 0.2, \quad 
\alpha_4 = 0.1, \quad 
\alpha_5 = 0.5.
\]
According to our classification, individuals 1, 2, and 5 are socialists, while individuals 3 and 4 are liberals.  
A stable configuration for this system is:
\[
p_1 = p_2 = 0, \quad p_3 = 2, \quad p_4 = 3, \quad p_5 = 10.
\]
Total production equals \( 15 \), yielding \( P = 3 \). Thus:
\[
\begin{aligned}
c(p_1, P) &= 0.9, &\quad c(p_2, P) &= 0.9, &\quad c(p_3, P) &= 2.3,\\
c(p_4, P) &= 3, &\quad c(p_5, P) &= 7.9.
\end{aligned}
\]
The total distributed amount equals total production, confirming internal consistency. However, not all individuals receive what they consider a just distribution:
\[
\begin{aligned}
c_1(p_1, P) &= 1.2, &\quad c_2(p_2, P) &= 1.8, &\quad c_3(p_3, P) &= 2.2,\\
c_4(p_4, P) &= 3, &\quad c_5(p_5, P) &= 6.5.
\end{aligned}
\]

\begin{table}[h!]
\centering
\renewcommand{\arraystretch}{1.2}
\begin{tabular}{c|ccccc}
 & $I_1$ & $I_2$ & $I_3$ & $I_4$ & $I_5$ \\
$\alpha_i$ & 0.4 & 0.6 & 0.2 & 0.1 & 0.5 \\
$a_i$ & 1 & 2 & 2 & 4 & 10 \\
$p_i(t)$ & 0 & 0 & 2 & 3 & 10 \\
$P$ & 3 & 3 & 3 & 3 & 3 \\
$c_i(t)$ & 1.2 & 1.8 & 2.2 & 3 & 6.5 \\
$c(p_i, P)$ & 0.9 & 0.9 & 2.3 & 3 & 7.9 \\
\end{tabular}

{Stable configuration of the system in the example.}
\label{tab:stable_system}
\end{table}

This is the only stable point, assuming that the fifth individual—being a socialist with significantly above-average capacity—produces at full capacity. In that case, total production is at least \( 10 \), implying \( P \geq 2 \). Since \( a_1 = 1 < P \), we have \( p_1 = 0 \). Individuals 3 and 4, being liberals, produce either at their maximum or at the level \( P \), ensuring total production exceeds \( 10 \), thus \( P > 2 \). Consequently, \( a_2 = 2 < P \) implies \( p_2 = 0 \). The third individual (a liberal) produces \( p_3 = 2 \), and the fourth (also a liberal) works at \( p_4 = P \). Because total production cannot exceed \( 16 \), we have \( P \leq 3.2 < a_4 \), and hence \( p_4 = P \). This yields:
\[
0 + 0 + 2 + P + 10 = 5P 
\quad \Rightarrow \quad 
12 = 4P 
\quad \Rightarrow \quad 
P = 3.
\]

Alternatively, if all individuals produce at a common level not greater than the minimal capacity of individuals in the society (not greater than 1), the system would also remain stable, but this stability is trivial.

The point toward which the system converges from a non-stable state depends on the initial state of the system.

\section{Dynamics of a Simple Taxation Model}

While the stable positions are relatively easy to identify—since, in a stable state, socialists either produce at full capacity, or the average, or not at all, and the liberals’ level of production is entirely determined by that of the socialists—the dynamics leading to those equilibria can exhibit interesting behavior.

For example, if at the initial stage no socialist produces above the average, usually, all of them will eventually reduce their productivity to zero\footnote{Under very special choice of the parameters, it may happen that in in time the average decrease below what one of the socialists produces, and this will change the outcome}. Consequently, liberals will also progressively reduce their productivity converging to zero.

On the other hand, in the example discussed previously, if all individuals start by producing at full capacity, the fifth individual will continue to do so, since she is a socialist producing above average. As this implies that the average production is at least \( 2 \), the third individual, being a liberal producing below average, will also produce at full capacity. In contrast, the first and second individuals will produce less than in the previous step, as their productivity decreases by at least \( 0.1 \) per iteration. This decline reduces both total and average production, and within at most twenty iterations their production becomes \( 0 \). The fourth individual, being a liberal producing above average, will also decrease her production. However, she will not reduce it to the average level, and since the average continues to decrease with each iteration, we have
\[
p_4(t) \to 3,
\]
although this limit is never actually reached.

Thus, we observe that in the dynamic evolution of productive systems, the outcomes tend to converge toward one of the stable points identified previously. It is reasonable to conjecture that this convergence behavior holds more generally\footnote{If we introduce another socialist individual with capacity slightly above \( 3 \), her behavior will depend on the size of \( \alpha_6 - \alpha \). If this difference is small, the decrease in her productivity will be slow, and eventually the average production will fall below her productivity. After that, she will begin increasing her production toward full capacity, and the stable point will lie slightly above \( 3 \). Conversely, if \( \alpha_6 - \alpha \) is large, her productivity may fall faster than the average, leading her to remain inactive after a few iterations. In that case, the average productivity will remain below \( 3 \).}.

\begin{rem}
Another observation concerning the example in the previous section is that changes in the level of taxation affect total production, which may either increase or decrease. For instance, if \( \alpha = 0.41 \), the first individual becomes a liberal and begins producing at full capacity, increasing total production by \( 1 \). However, if \( \alpha = 0.51 \), the fifth individual becomes a liberal and reduces her productivity to the average level. This results in a system where the only stable states correspond to all individuals producing the same amount, which must lie within the interval \( [0,1] \). Consequently, total productivity does not exceed \( 5 \), which is significantly lower than in the case \( \alpha = 0.3 \).
\end{rem}

\section{How Realistic Is the Simple Taxation Model?}

Although highly simplified, the model demonstrates sufficiently complex behavior. The assumption that part of the total production is redistributed corresponds well to modern flat-taxation systems, where a fixed proportion of individual production (in monetary form) is collected as taxes and spent on public goods. Each individual then receives an equal share of the public goods.

We clearly do not address the fact that different individuals may derive different levels of benefit from the public goods. Nor do we account for the fact that productivity may be influenced not only by satisfaction with the distribution system but also by intrinsic motivation or the positive emotions associated with productive activity. These simplifications, however, seem reasonable as a first approximation.

Nevertheless, the model exhibits effects that resemble phenomena observed in countries with extensive welfare systems. Some individuals withdraw from contributing to total production even when capable of doing so, while others with similar capacities continue contributing. Moreover, individuals with high productive capacity may choose to reduce their contributions in so called income targeting \cite{CP,TT}.

We also omit the possibility that individuals form opinions not only about how redistribution affects themselves but also about how it affects others. A model incorporating such “envy” effects would be considerably more complex, yet it might better capture the behavior of modern progressive tax systems. One possible extension would be to describe each individual’s stance with two parameters: one governing taxation for those producing less than or equal to themselves, and another for those producing more. This could serve as a promising direction for future research.

Another simplifying assumption in the present model is that each individual evaluates the value of their own work in a universal and perfectly informed manner, having full knowledge of total production. Developing a model that accounts for incomplete information about individual and collective productivity would be a valuable direction for further study.

\end{document}